\documentclass[a4paper,fleqn,usenatbib]{mnras}

\usepackage{newtxtext,newtxmath}

\usepackage[T1]{fontenc}
\usepackage{ae,aecompl}

\usepackage{graphicx}	
\usepackage{amsmath}	
\usepackage{amssymb}	

\title[M31 fully reshaped by a 2-3 Gyr old merger]{A 2-3 billion year old major merger paradigm for the Andromeda galaxy and its outskirts}

\author[F. Hammer et al.]{
Francois Hammer$^{1}$\thanks{E-mail: francois.hammer@obspm.fr},
Y. B. Yang$^{1}$,
J. L. Wang$^{1,2}$,
R. Ibata$^3$,
H. Flores$^{1}$,
and M. Puech$^{1}$
\\
$^{1}$GEPI, Observatoire de Paris, PSL Research University, CNRS, Place Jules Janssen, 92190 Meudon, France.\\
$^{2}$NAOC, Chinese Academy of Sciences, A20 Datun Road, 100012 Beijing, PR China.\\
$^{3}$Observatoire Astronomique, Universit\'{e} de Strasbourg,
CNRS, 11, rue de l Universit\'{e}, F-67000 Strasbourg, France.
}

\date{Accepted 2017 December 19. Received 2017 November 06; in original form 2017 September 07}

\pubyear{2017}

\begin{document}
\label{firstpage}
\pagerange{\pageref{firstpage}--\pageref{lastpage}}
\maketitle

\begin{abstract}
 Recent observations of our neighbouring galaxy M31 have revealed that its disk was shaped by widespread events. The evidence for this includes the high dispersion ($V/\sigma$ $\le$ 3) of stars older than 2~Gyr, and a global star formation episode, 2--4~Gyr ago. Using the modern hydrodynamical code, GIZMO, we have performed 300 high-resolution simulations to explore the extent to which these observed properties can be explained by a single merger. We find that the observed M31 disk resembles models having experienced a 4:1 merger, in which the nuclei coalesced 1.8-3 Gyr ago, and where the first passage took place 7 to 10~Gyr ago at a large pericentre distance (32~kpc). We also show that within a family of orbital parameters,  the Giant Stream (GS) can be formed with various merger mass-ratios, from 2:1 to 300:1. A recent major merger may be the only way to create the very unusual age-dispersion relation in the disk. It reproduces and explains the long-lived 10~kpc ring, the widespread and recent star formation event, the absence of a remnant of the GS progenitor, the apparent complexity of the 3D spatial distribution of the GS, the NE and G Clumps and their formation process, and the observed slope of the halo profile. These modelling successes lead us to propose that the bulk of the substructure in the M31 halo, as well as the complexity of the inner galaxy, may be attributable to a single major interaction with a galaxy that has now fully coalesced with Andromeda.
\end{abstract}




\section{Introduction}
For a very long time, the M31 galaxy has been considered as a Milky Way (MW) twin since both galaxies share a similar morphological classification (Sb and Sbc, respectively), possess a large bar in their centre, and have a prominent galactic disk. However, recent large surveys (PAndAS: \citealt{Ibata2014}, PHAT: \citealt{Gilbert2009}, SPLASH: \citealt{Dalcanton2012}) have provided an enormous wealth of data probing the M31 halo, disk and bulge, respectively.  The ensuing findings emphasise how different the two galaxies that dominate the Local Group are:
\begin{enumerate}
\item The classical bulge of M31 is much more prominent than the pseudo-bulge of the MW \citep{Courteau2011};
\item  The M31 galaxy possesses a massive ring of star formation at 10~kpc, which is especially clear in the mid-IR \citep{Barmby2006,Gordon2006};
\item While only $\sim$ 10\% of the MW's stars are in a thick disk \citep{Juric2008}, almost the whole M31 disk is thick and presents a far steeper stellar age-velocity dispersion relation \citep{Dorman2015};
\item The M31 disk within 20 kpc shows a global, prominent star formation episode, 2 to 4 Gyr ago \citep{Bernard2015a,Williams2015};
\item The M31 halo is simply exceptional by the variety of streams \citep{Ibata2014}, and includes the Giant Stream (GS, \citealt{Ibata2001}) which dominates the star-counts at projected distances between $\sim 30$ and $120$~kpc.
\end{enumerate}
The MW is known to be a relatively quiet galaxy without any recent,  major merger event, while M31 has properties quite close to an average, representative local spiral galaxy  \citep{Hammer2007}. Calculations from the $\Lambda$CDM theory predict approximately one major merger in the past history of galaxies in the MW-M31 range of mass \citep{Stewart2008}, which is confirmed by observations (see \citealt{Rodrigues2017} for a recent update). What then could be the cause of such considerable differences in the past histories of these two giant galaxies which share the same environment?

Many previous studies have suggested an ancient major merger event in the past history of M31, based on its classical bulge \citep{Kormendy2013,Courteau2011} and its halo rich in streams of various metallicities \citep{vandenBergh2005}. However, they are not predictive about the epoch of the last merger, or  more precisely of the final coalescence of the two nuclei. There has been only a single attempt \citep{Hammer2010} to study a 5--8~Gyr old major merger reproducing the disk (stars \& \ion{H}{I}), bulge/total (B/T) luminosity ratio, and the 10 kpc ring, though it does not convincingly reproduce the GS that is likely a dynamically young structure given its sharpness and spatial contrast. A recent overview of all the known properties of M31 is perhaps missing, and we refer to the \citet{Davidge2012} study for a more comprehensive overview of the origins of M31 and past modelling efforts (see also \citealt{Ferguson2016} for the properties of the outskirts). Indeed, to date, most efforts have been devoted to reproducing individual features of M31 assuming a single minor merger event. Such studies have found that:
\begin{enumerate}
\item Assuming that the GS is a trailing tidal tail of a minor merger (few $10^{9} M_{\odot}$), N-body simulations are able to reproduce well the structure and kinematics of the GS and the NE and W shells together \citep{Fardal2006,Fardal2007,Fardal2008,Fardal2013,Mori2008,Kirihara2014,Kirihara2017, Sadoun2014}. Such models favour a small spiral galaxy structure for the progenitor \citep{Fardal2013,Kirihara2017} and an event duration of $\sim$ 1 Gyr. Finding evidence for the (still undiscovered) existence of the remnant of the GS progenitor dwarf galaxy would constitute a major success for this minor merger model.
\item The  inner and 10 kpc rings have been proposed as having been caused by an interaction with M32 \citep{Block2006,Dierickx2014}; however, such a passage predicts an expanding ring while the 10 kpc ring appears to have been stable for at least 500 Myr \citep{Lewis2015}.
\item \cite{Bernard2012} suggested an interaction with M33 on the basis of the coincidence of a shared star formation episode 2-4 Gyr, while \cite{McConnachie2009} came to a similar conclusion based on the presence of tidal disturbances that resemble tidal tails around M33. However, the discovery that the whole of the M31 disk up to 20 kpc experienced a burst of star formation 2-4 Gyr ago renders it less plausible to be provoked by a single passage of a satellite of $\sim$ 10\% of the mass of M31 \citep{Williams2015}. 
\end{enumerate}
\citet{Tanaka2010} concluded their study of the numerous streams in the M31 halo by noticing that to explain them all, it would require $\sim$ 15 accretions of sub-haloes with masses of typical dwarf galaxies. Formerly, the main argument in favour of minor mergers was to ensure the survival of the M31 disk. However, it has been shown that disks may be rebuilt after a sufficiently gas-rich major collision \citep{Hammer2005,Hammer2009,Hopkins2009}. Furthermore,  the spiral arms in M31 seem not to be triggered by the classical density wave theory \citep{Tenjes2017}, which distinguishes it even more from a quiescent galaxy. It is therefore now necessary to reevaluate the plausibility of a single major merger versus numerous very minor events as a means to reproduce most of the exceptional properties of M31 listed above. Which event(s) is (are) responsible of most of these structures, from the bulge to the halo? Could they be reproduced by a gas-rich major merger? This latter possibility has to be consistent with the "dynamically-young" structures in the halo such as the GS and shells, and account for recent, widespread events in the disk that are suggestive of a recent merger episode.


Fully exploring the major merger paradigm is clearly a daunting task. Its study is far more complex and time consuming than that of minor mergers, since:
\begin{enumerate}
\item Models of very minor mergers have only to complete one orbit of the progenitor, and do not affect the overall structure of the galaxy;
\item Models of major or intermediate mergers have to be evolved over several orbits until the complete destruction of the original disks;
\item An extraordinary large space of parameters has to be investigated to reproduce the very numerous properties of the resultant galaxy; 
\item Realisations of the Giant Stream at the resolution of PAndAS require $\sim$ 200 k particles in a minor merger (see, e.g., \citealt{Kirihara2017}), while we estimate it requires 30-40M particles for a major merger model.
\end{enumerate}
Notwithstanding the above, the aim of this paper is to verify to which extent the Andromeda (and its outskirts) numerous structures can be reproduced by a unique merger. The major constraints are provided by:
\begin{enumerate}
\item The kinematics of the M31 disk from its rotation curve and its age-velocity dispersion relation;
\item The structural parameters of the inner M31 galaxy: B/T, bar, structure of the 10~kpc star-forming ring, bulge, bar and disk size, and \ion{H}{I} disk;
\item The structures surrounding the thick disk, including the warp (e.g., such as the NE Clump, the Northern Spur, the Warp \& the G1 Clump), the NE and W shells and the GS;
\item Stellar ages and metal abundances of the above structures.
\end{enumerate}
Ideally such a model, if it exists, should be able to reproduce the many detailed properties of the M31 and outskirts structures. 
However, we are aware that aiming to reproduce in details the spatial and kinematic structure of a nearby galaxy with a major merger study is perhaps inextricably challenging, because:
\begin{enumerate}
\item The accuracy of the calculations are limited for numerical reasons, especially because of the huge contrast in the required number of particles to explain simultaneously the central galactic regions and the faint streams in the outskirts;
\item They are further limited by the first pericentres that cannot be estimated at much better accuracy than $\sim$4\%  (see \S2);
\item Important features, e.g., shells (see \citealt{Cooper2011}) are evolving with time, meaning that they can be found like they are in M31 but perhaps not simultaneously with other features;
\item A major merger may also produce additional features that could be a consequence of the --- completely unknown --- internal mass distribution in the progenitors.
\end{enumerate}
In this paper we propose a first attempt in modelling the huge wealth of information recently obtained about our closest giant neighbour, the Andromeda galaxy. Besides trying to reproduce most of the observed properties together, we aim at proposing a physical interpretation of as many morphological details as we can, in the frame of a single event. In Section 2 we establish the framework of the simulations, including the hydrodynamics, the star-formation implementation and initial conditions, as well as the expected limitations when simulating M31. Section 3 presents the results and compares them to the observations, and then Section 4 discusses whether M31 could be or not the result of a single major event instead of numerous minor mergers.

\section{Simulations}

The faintness of several features (e.g., the streams)  requires a very large number of particles to model correctly. This limits us to only being able to reproduce the brightest halo features including the Giant Stream and perhaps suggesting further some possible mechanisms to explain, e.g., Streams A to D reported by the PAndAS team.  We start simulations using similar orbits than those used in \citet{Hammer2010}, most orbital parameters requiring only fine-tuning at the level of 10\% variations (see initial and adopted parameter ranges in Table~\ref{tbICs}). However the 2--4 Gyr star formation history in the whole disk \citep{Williams2015} and the steepness of the age-dispersion relation found by \citet{Dorman2015} cannot be reproduced by an ancient merger (coalescence of the nuclei 5 to 6 Gyr ago) as suggested in \citet{Hammer2010}. We indeed verify that such event is unambiguously followed by a star formation episode 5--6 Gyr ago and velocity dispersions far below the observed values. 
 
So next we investigate whether a more recent, 2--3 Gyr old coalescence of the nuclei, could explain most of the widespread activity in the disk. Since observations are very constraining, and hence very demanding in terms of controlling the physics, we considerably improve our former, GADGET2 model by performing simulations with the hydrodynamical resolver GIZMO. A significant part of the study has been devoted to exploring a limited number of parameters, firstly the mass ratio (mr) and the pericentre. Most features depend considerably on these parameters, since the former defines the number of passages (the smaller secondary mass, the larger the number of passages) and the latter provides the time elapsed between two passages. Then we have had to search for optimising resonances that are revealed by the presence of the bar, and a considerable amount of work was devoted to finding parameters that amplify them mostly by varying the progenitor sizes. Besides this we also reproduce the bulge, disk scale-length and rotation curve (see Appendix A) as in \citet{Hammer2010}. 


\subsection{Hydrodynamical resolver and star formation and feedback implementation}
\subsubsection{GIZMO: an optimised hydrodynamical solver}
GIZMO is a recently published code for N-body hydrodynamical computation that has been developed by \citet{Hopkins2013,Hopkins2014}, to whom we refer for a full description of the algorithm. This code is developed and optimised from GADGET3, which is an worldwide-used N-body/SPH code by \citet{Springel2005}. In addition to all the advantages of GADGET3, \citet{Hopkins2013} introduces accurate (to 2nd order) hydrodynamical solvers that rely on Lagrangian numerical methods, i.e., the mesh-less finite-mass method (MFM) and the mesh-less finite-volume method (MFV). This makes GIZMO able to perform massive parallel computations, and provides very accurate solutions of gas hydrodynamics, as well as the best conservation of mass, momentum and energy. These advances motivated us to change our simulation software from GADGET2 to GIZMO.  Simulations presented in this paper use the MFM hydrodynamical solver and a fixed softening 0.16~kpc for a 2-million particle simulation, and 0.08~kpc for 20-million particle simulation, respectively.
	
\subsubsection{Star formation and feedback implementation}	
Following Cox et al. (2006), we have successfully implemented a simplified star formation and feedback model into GIZMO, applying the same methodology as done by Wang et al. (2012). This model can describe well the star formation history over large scales in galaxies and has been used intensively in the last decade (see, e.g., \citealt{Hammer2010,Wang2012,Wang2015}).

In high gas density regions, gas can form stars. The star formation rate of each gas particle is calculated according to the local gas density and the local dynamic time based on the observed Kennicutt-Schmidt law \citep{Kennicutt1998}. A stochastic method \citep{Springel2000,Cox2006} is used to convert gas particles into star particles.

During star formation, feedback processes are important to regulate star formation. In this method, the energy released by supernovae during star formation is first stored in a new reservoir of internal energy, which provides additional pressure to support the gas and prevent it from further collapsing to form stars. This feedback energy can be thermalised by a free parameter, which controls the timescale of this thermalisation and then the feedback strength.

The radiative cooling processes are necessary for gas to cool down and collapse to form stars. In this work, the cooling rate is calculated with the method implemented in GIZMO, which treats the gas as a primordial plasma, and the ionisation states of H and He for a collisional ionisation equilibrium are assumed \citep{Katz1996}.

\subsection{Initial conditions, parameter choices, and limitations}

  \begin{table*}[!ht]
    \caption{Initial and Adopted Conditions for a Major Merger Model for M31.}
    \begin{center}
      \begin{tabular}{ccllll}
        \hline\hline
        Ingredient & Tested range & Comments  & Adopted Range \\
        \hline
        total mass & 8.25$\times$$10^{11}M_{\odot}$ & 20\% of baryons & -\\
        mass ratio & 2-5 & to reform B/T$\sim$0.3 &  4.0 (3.5-4.25)\\
        $f_{gas}$ Gal1 & 0.4-0.6 &  expected at z=1.5$^{(a)}$  & 0.4-0.6\\
        $f_{gas}$ Gal2 & 0.6-0.8 &  expected at z=1.5 & 0.6-0.8\\
        Orbit & near polar & to form the ring &  - \\
        Gal1 $\theta^\prime$$^{(b)}$ & 65 to 100 & Giant Stream  & 35-75\\
        Gal2 $\theta^\prime$$^{(b)}$ & -50 to -70 & Giant Stream  & -60 to - 70 \\
        Gal1 $\phi^\prime$$^{(c)}$ & 115 to 175 & Giant Stream & 165 \\
        Gal2 $\phi^\prime$$^{(c)}$ & 75 to 110 & Giant Stream  & 95-105\\
        $r_{pericentre}$ & 28-40 kpc & see the text &  32 kpc (31-33 kpc)\\
        Feedback & 1-5$\times$median$^{(d)}$ & to preserve gas  & 1-2.5$\times$median$^{d}$\\
       \hline
      \end{tabular}
      \\
      $(^{a})${\citet{Rodrigues2012} found $f_{gas}$=0.5-0.65 in galaxies with $M_{baryon}$=0.8-2.2 $10^{11}M_{\odot}$ at z=1.5.}\\
      $(^{b})${Rotation along the $y^\prime$ axis.}\\
      $(^{c})${Rotation along the $z^\prime$ axis.}\\
      $(^{d})${In few simulations, the feedback is assumed to be high before coalescence of the nuclei and later on, assumed to drop to the medium or low feedback values of \citet{Cox2006}.}
    \end{center}
    \label{tbICs}
  	\end{table*}

We follow the same method used in \citet{Hammer2010} \citep[see also][]{Wang2012} to create initial conditions and to test the stability of the progenitors in isolation. Each progenitor is assumed to possess only 2 components: a dark matter halo and a thin disk that includes stars and gas. We define the gas fraction of the progenitor to be the fraction of gas mass to the total baryonic mass and the mass ratio to be the ratio of the total mass of each galaxy. The density distributions of each component and their N-body realisations follow exactly the method described in \citet{Barnes2002}\footnote{See also http://www.ifa.hawaii.edu/$\sim$barnes/software.html.}. Both the stellar disk and the gas disk are assumed to be thin with a scale-height equal to 1/10 of the scale-length. The scale-length of the gas disk is assumed to be 3 times larger than that of the stellar disk \citep{Cox2006}. The scale of the dark-matter halo core ($a_{\rm halo}$) is chosen to ensure that the initial disk is consistent with the Tully-Fisher relation. For the simulations analysed in this paper $a_{\rm halo}$ are 11 and 4 kpc, for the large and small progenitor, respectively.

Most of the parameters for initial conditions used in this work are listed in Table~\ref{tbICs}. We use the same dark matter fraction (20\%) and density profile as in \citet{Hammer2010}.  Half of our simulations were run with 2 million particles. To test for convergence, we also ran a simulation with 20 million particles. We found the resulting end-point structures were qualitatively unchanged, which indicates that convergence has been achieved (see Appendix B). 

 Each simulation was launched on a parabolic orbit $\sim$ 1 Gyr before the first passage to allow sufficient relaxation of the progenitors. The parameters in Table~\ref{Models} are defined in the simulation frame where the orbital plane of the merger is always put in the $x^\prime$-$y^\prime$ plane with the orbital spin aligned with the $z^\prime$-axis; the initial pericentre is always on the positive side on the $x^\prime$-axis. To provide initial conditions, the progenitor disks are first put in the $x^\prime$-$y^\prime$ plane with their rotational spin along the $z^\prime$-axis, then rotated about the axes by angles $\theta^\prime$ and $\phi^\prime$ (for a spherical coordinate system) that are given in Table~\ref{Models}.

\subsubsection{Constraints on the mass ratio and pericentre from the HI disk}
\begin{figure}
	\includegraphics[width=\columnwidth]{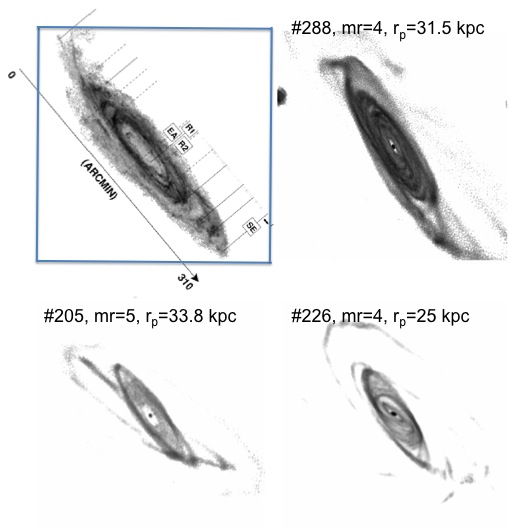}
    \caption{Comparison of the \ion{H}{I} disks provided by models with the same orbital parameters but for which the pericentre (right panels) varies from 25 to 32 kpc or the mass ratio from mr=4 to 5. {\it From top to bottom, left to right:}  Observations from \citealt{Chemin2009} and 3 models for which number, mass ratio and pericentre are indicated on the top of each panel;  the size of the blue box is 60 kpc, and all plots are on the same scale.}
    \label{HI}
\end{figure}

The Andromeda galaxy is characterised by a prominent classical bulge and a significant bar \citep{Athanassoula2006,Beaton2007}.  Based on its polar orbit, the previous model of a 3:1 merger by \citet{Hammer2010} also aimed to reproduce its star-forming, 10 kpc ring, as well as its HI disk (see their Fig. 6). Here we have to also account for the $\gtrsim$ 2~Gyr old wide perturbations of the M31 disk. We identify the \ion{H}{I} disk to be the feature that is least stable against either a recent merger or a significant amount of feedback. Figure~\ref{HI} shows how varying the merger mass ratio and pericentre affects the \ion{H}{I} disk size. It suggests a merger mass ratio (mr) from 3 to 4 and a pericentre larger than $\sim$ 32 kpc. Indeed a large pericentre brings more angular momentum to the remnant disk. From our examination of our whole set of models (see also Figure~\ref{HI}) we found that mr$\ge$ 4.5 as well as $r_p<30$~kpc are likely excluded values to reproduce the M31 \ion{H}{I} disk size.


\begin{figure}
\centering
\includegraphics[width=8.0cm]{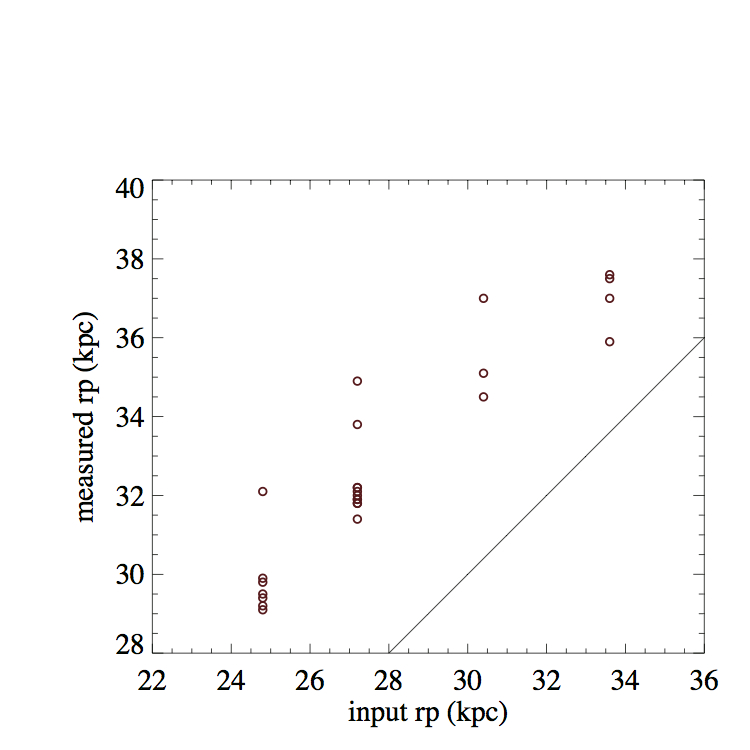}
\caption{Input versus actual pericentre in the simulations. The dispersion in measured pericentre is due to different initial conditions that affect the mix between stars and gas.}
\label{figrp}
\end{figure}

\subsubsection{Accuracy of the pericentre and its adopted choice}
\label{limitations}
We found that important properties of a merger remnant a few Gyr after  the coalescence of the nuclei are well correlated with the pericentre distance at the first passage. Even small changes of the first pericentre, e.g., of 1.2 kpc (i.e., 4\% of 32 kpc), will change the elapsed time between first and second passages by $\sim$ 1 Gyr, affecting the final  coalescence time, and then the timing of the epochs of star formation in the rebuilt disk. However, the actual pericentre value is not easy to control just from the initial pericentre because it also depends on the different properties of the initial progenitors, such as the gas fraction, size of both stellar and gas disks, and star formation history. Different initial conditions (ICs) will lead to different star formation histories and different mass distributions, resulting in changes in, e.g., the mass ratio at later passages, and then to strong variations in dynamical friction. This is demonstrated in Figure~\ref{figrp} in which we show various simulations for which several values of feedback, initial gas fraction, and initial disk sizes have been implemented. We find a strong trend that almost follows the behaviour of a point mass (shown as a straight line in Figure~\ref{figrp}), and differences with point mass can be easily corrected for. Besides this, we find a scatter of $\sim$ 10\% solely due to the physical differences in the progenitors (see points in Figure~\ref{figrp}). After realising this, we analysed each simulation and relaunched it with slight changes of the initial position and velocity at the beginning of the simulation, in order to have to the simulation run within 4\% of the desired pericentre. The pericentre value is found to be a key parameter in our simulations for reproducing as accurately as possible the properties of M31. 

We found that adopting a value of $r_p \sim 32$ kpc for the first pericentre distance can give rise to final simulation structures in good qualitative agreement with the observations, as we show below. However the large M31 disk size (especially the \ion{H}{I} disk) may favour even higher values. Increasing the pericentre significantly would have been problematic, as it would increase the absolute uncertainty (in kpc) on the pericentre, and then on the elapsed time between, e.g., first and second passages. In such conditions, it would then have been difficult to compare different simulations for exploring the parameter space, and even to reproduce the same result with two simulations at different resolutions (e.g., we need to increase resolution to reproduce the faint features). Future work will explore further the consequences other choices of the $r_p$ parameter.

\subsubsection{Optimising the parameters to reproduce main features: the bar and the 10kpc ring, disks, bulge and rotation curve}

\begin{figure}
	\includegraphics[width=\columnwidth]{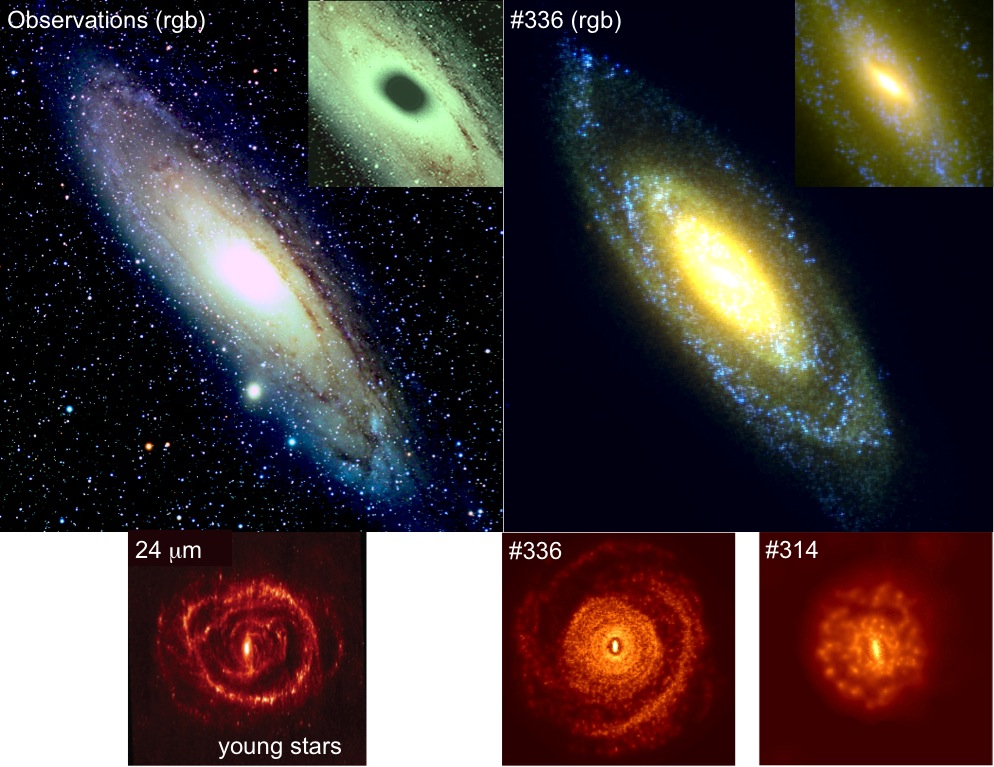}
    \caption{{\it Top row:} Comparison of red-green-blue (rgb) composite-colour images of the M31 disk (left) with our highest resolution simulation (right, model 336, see Table~\ref{Models}). The spectral energy distribution of each particle was calculated using the \citealt{Bruzual2003} population synthesis model, assuming a single stellar population and seven metallicity tracks. It does not account for dust, explaining part of the differences between images (see, e.g., dust lanes in the observed disk). On the top-right of the observed image, an insert (a 26.6 kpc wide box) shows BVRz observations and the superposed central bar (K-band, in black) by \citealt{Beaton2007}, which can compared to the simulation for which we have improved the contrast in the central region (see insert on the top-left of the simulated disk). {\it Bottom row:} Comparison of the star forming regions in the M31 disk within a 40~kpc wide box. From left to right,  the 24 $\mu$m (\citealt{Gordon2006}) reconstructed face-on view of the bar and of the 10~kpc ring, and simulations of $<$ 0.2 Gyr old stars for models 314 and 336 (see Table~\ref{Models}).}
    \label{barring}
\end{figure}

For each set of orbital parameters (see Table~\ref{tbICs}) we changed the scale-lengths of the initial stellar and gas disks in order to optimise the bar and the 10 kpc ring. The presence of the bar is also supported by the kinematics of the central regions \citep{Opitsch2017}. Figure~\ref{barring} shows that we succeed in reproducing a strong bar accompanied by a young star distribution similar to the observed one. It also demonstrates that the orientation of the bar is distinct from that of the major axis of the disk, as is observed (see, e.g., \citealt{Athanassoula2006} and compare the inserts in the top panels of Figure~\ref{barring}).  However our bar realisations often lead to thinner bars than that observed. The 10 kpc ring is often found off-centred (see bottom-left panel of Figure~\ref{barring}) as in the observations \citep{Lewis2015}, and our modelling may be consistent with the (observed) presence of an outer ring (or a part of it, see the bottom-middle panel of Figure~\ref{barring})

Besides the need to reproduce the actual properties of M31, an important motivation for favouring models that produce bars is that M31 possesses a relatively concentrated density profile, with the bulge representing 30\% or more of the total stellar mass (see, e.g., \citealt{Blana2017}). As shown in \citet{Wang2012}, the properties/spatial distributions of the streams strongly depend on the mass concentration of the remnant galaxy. We have also used intensively the \citet{Amorisco2015} study of the dynamics of tidal streams to optimise the reproduction of the properties of the Giant Stream (see Sect.~\ref{outskirts}). We have selected 5 fiducial models, whose parameters (see Table~\ref{Models}) show a particularly narrow range. They were selected after a considerable search effort, including $\sim$ 300 simulations, half with 500k particles and half with 2M particles. 

As in \citet{Hammer2010}, we reproduce the \ion{H}{I} disk (compare Figure~\ref{HI} with their Fig. 6), the stellar density profile of the disk, the bulge to disk ratio as well as the rotation curve (see, e.g., their Fig. 5 and Appendix A of this paper).  This is because for a gas rich merger, the disk settles within $\sim$ 1 Gyr after the coalescence of the nuclei, then similar orbits with similar mass ratios provide similar disk scales and rotation.  Our simulations produce a relatively large scale-length final disk (3.3, 5.8, 3.5, 3.9, and 4.8 kpc for models 255, 276, 288, 290 and 314, respectively) although often smaller than the M31 disk value (5.8 kpc, see discussion in \citealt{Hammer2007} and also \citealt{Blana2017}). The disk scale-length of M31 (as well as the \ion{H}{I} disk size) is particularly large when compared to those of similar spiral galaxies (see, e.g., \citealt{Hammer2007}), and if it results from a merger, a large pericentre would naturally form a large remnant disk. In the previous section, we discussed the difficulty in studying large pericentres: this is likely the main limitation of our study. 

\begin{table*}
	\centering
	\caption{Initial parameters of the 5 fiducial models plus that with 20 million particles (model 336). The second column indicates the number of particles (M=$10^6$). The third and fourth columns indicate the pericentre and the mass ratio. The fifth column gives the feedback prescription (i.e., versus the median value of \citealt{Cox2006}), and in Model 314 and 336 we have let vary the feedback value before and after coalescence of the nuclei (first and second value, respectively). The next four columns provide the angle (in degrees) of the progenitors against the orbital plane (see text) and $1^{st}$ and $2^{nd}$ indicate primary and secondary progenitors, respectively. The last columns indicate the initial disk scale length (in kpc) of stars and gas in the progenitors, and their initial gas fraction, respectively.}
	\begin{tabular}{lcccccccccccccr} 
		\hline
		Model & N particles & $r_p$ & mr & Feedback & $\theta^\prime$ & $\phi^\prime$ & $\theta^\prime$ & $\phi^\prime$ & $h_s$ & $h_g$ & $f_{gas}$ & $h_s$ & $h_g$ &$f_{gas}$\\
		 & & kpc & & $\times$ Cox median & $1^{st}$ & $1^{st}$ & $2^{nd}$ & $2^{nd}$ & $1^{st}$ & $1^{st}$ & $1^{st}$ & $2^{nd}$ & $2^{nd}$ & $2^{nd}$  \\
		\hline
255  &   2M   & 29.4 & 4.0 & 1 & 70 & 165 & -60 & 85  & 2.8 & 8.4 & 0.4 & 2.8  &  8.4  & 0.6\\ 
276   & 2M &  31.9  & 4.0  & 1 & 70 & 155 & -60  & 85  &  2.8  & 16.8 & 0.6 & 2.8  & 16.8 &  0.6\\  
288   & 2M  & 31.5  & 4.0  & 1.25 & 70 & 165 & -60  & 85   &  2.8 & 12.6 & 0.6 & 2.8  & 12.6  & 0.6\\
290   & 2M  & 32.0  & 4.0 & 1 & 70 & 165 & -60 & 105  & 2.8 &  8.4 & 0.4 & 2.8  &  8.4 &  0.6\\
314   & 2M  & 34.8 & 4.0 & 2.5-1 & 75 & 165 & -60 & 105  &   2.8  & 8.4 & 0.4 & 2.8  &  8.4  & 0.6\\
336  & 20M  & 32.8 & 4.0 & 2.5-1 & 70 & 165 & -60 & 100 & 2.8  & 8.4 & 0.5 & 2.8  & 8.4  & 0.8\\
		\hline
	\end{tabular}
\label{Models}
\end{table*}

\section{Results}
\subsection{The unusual properties of the M31 disk: stellar ages and kinematics}
\subsubsection{A 2--4 Gyr strong star forming event}
In M31, the outer-disk stars are moderately old or young and  \citet{Bernard2015b} found that most of them formed $\le$ 8 Gyr ago, while a quarter of them formed in the last 5 Gyr. Indeed, such an apparent rejuvenation of the disk is also observed in many spirals with similar mass and type as M31 (see, e.g., \citealt{GonzalezDelgado2017}). \citet{Williams2015} have undertaken a systematic study of the age of the stellar disk by sampling PHAT fields for which dust extinction is very small. They however limit their study to the sole period of the last 5 Gyr, which contains the 2--4 Gyr old strong burst of star formation (see also \citealt{Bernard2015a}). 

\begin{figure}
	\includegraphics[width=\columnwidth]{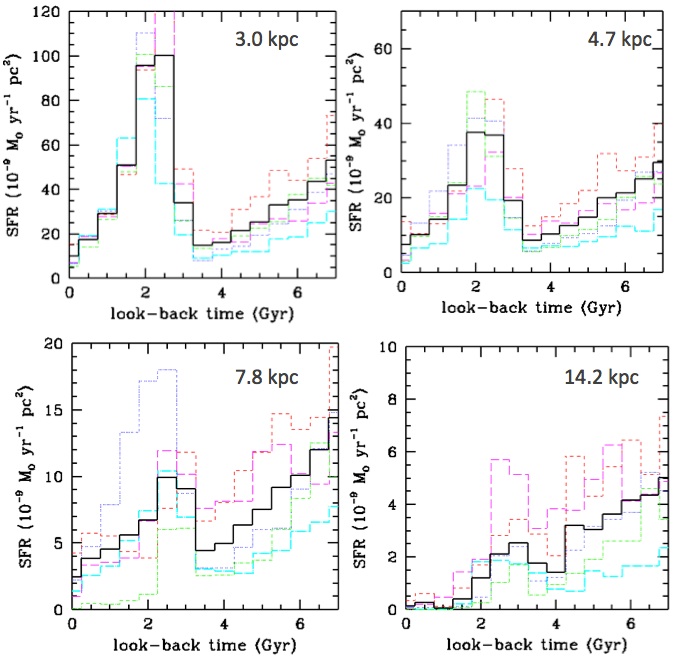}
    \caption{Star formation history from the five fiducial models of Table~\ref{Models}, as measured at 3, 4.7, 7.8 and 14.2 kpc along the disk, i.e., following fields studied by \citealt{Williams2015}. Green (dot - short dash), magenta (long dash), red (short dash), blue (dot), and cyan (dot - long dash) lines represent models 255, 276, 288, 290, and 314, respectively. The coalescence epoch is 3.0, 2.1, 2.1, 2.1, 1.8 Gyr ago for the later models, respectively. The black bold (solid) line represents the average of the later models, displaying the presence of a $\sim$ 2--4 Gyr old star formation episode over the whole disk.}
    \label{DISK_ages}
\end{figure}

\begin{figure}
	\includegraphics[width=\columnwidth]{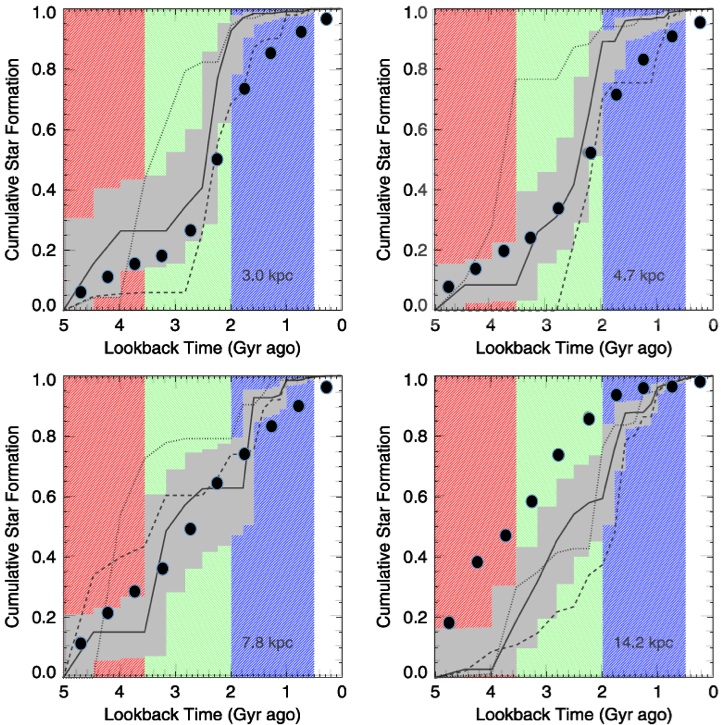}
    \caption{Same as Fig.~\ref{DISK_ages} (solid dots are averaged values of the five models) but for cumulative star formation to enable a direct comparison to Fig. 4 of \citealt{Williams2015}.}
    \label{DISK_cumul_ages}
\end{figure}

Figure~\ref{DISK_ages} presents the result of the 4:1 merger models for which  the coalescence of the nuclei occurred from 1.8 to 3 Gyr ago. All models show a significant burst of recent to moderately recent star formation, qualitatively similar to the observations. To compare in a more quantitative way, Figure~\ref{DISK_cumul_ages} reproduces Fig. 4 of \citealt{Williams2015} to which we have added the average value of the five models of Figure~\ref{DISK_ages}. Our simulations match quite closely the observations in the 3 first age panels, when accounting for the observational uncertainties (shown in grey), and for the model differences (see the dashed and dotted lines, and \citealt{Williams2015} for more details). Besides this, the burst of star formation seems too diluted in the outer disk field to reproduce quantitatively the observations, except for model 276 (see magenta, long-dashed line in the 4th panel of Figure~\ref{DISK_ages}), which shows the largest gas-disk scale-length. We interpret this as possibly due to our limitations in performing simulations with very large pericentres, i.e., those producing the largest disk scale-lengths.


\subsubsection{All stars older than two Gyr are in a thick disk}
\begin{figure}
	\includegraphics[width=\columnwidth]{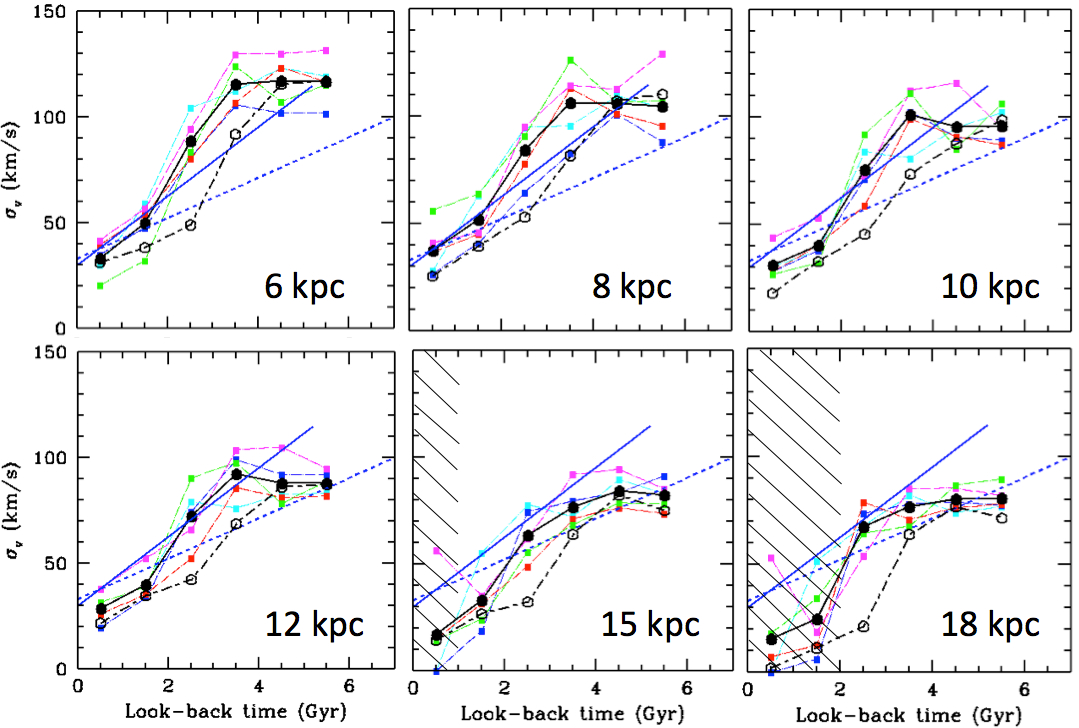}
    \caption{{\it From Top-Left to Bottom-Right:} Kinematic dispersion of the disk stars at 6, 8, 10, 12, 15 and 18 kpc. The (blue) solid and dashed straight lines are the best-fit to the data points at R$\ge$ 10 kpc with a constant SFR and decreasing SFR assumed by \citealt{Dorman2015}, respectively. Cyan, blue, red, magenta, and green represents models as in Fig.~\ref{DISK_ages}. The black bold (solid) line and solid dots represent the average of the models, while the bold dashed line and open dots represent the average value of the same models, but taken 1 Gyr later. The hatched areas in the two last panels indicate stellar ages for which the number of (young) stars are not sufficient to derive a reliable velocity dispersion. Note that the panels at 6 and at 8 kpc illustrate the Dorman et al. result that the kinematic dispersion of stars within 10 kpc can reach 120 km$s^{-1}$.}
    \label{starkin}
\end{figure}

The PHAT survey has allowed to establish an extinction map of the RGB stars in the M31 disk \citep{Dalcanton2015}, revealing a scale height (0.87 kpc) comparable to that of the MW thick disk. Only 10-15\% of the MW disk stars \citep{Juric2008} are in the thick disk, which is assumed to be formed 10--11 Gyr ago, and these stars are much more metal poor than those in the thin disk. 
This contrasts with the discovery of \citet{Dorman2015} that the  RGB stars in the M31 disk have large velocity dispersions (averaging to $\sim$ 90 km$s^{-1}$). It is a definitive proof of an overwhelming dominant thick ($V/\sigma \le$ 3) disk in M31. The M31 disk is much hotter than that of the MW. \citet{Dorman2015} also found a very steep age-velocity dispersion correlation. Figure~\ref{starkin} shows that such behaviour is expected for a merger remnant after a recent (2--3 Gyr ago)  coalescence of the nuclei, and the open points (and bold dashed lines) show that  elapsed times since coalescence larger than 3--4 Gyr would be too large to keep the disk as kinematically hot as is observed. In this Figure we have assumed an aperture (1 kpc) that corresponds to the smoothing radius used by \citet{Dorman2015} to create their 2D dispersion maps. Notice that the high $\sigma$ values cannot be caused by the initial velocity dispersion in the progenitors, which was set to 10 km$s^{-1}$. It is however closely associated to the highly retrograde motion of the primary in the orbital plane ($\phi^\prime$ slightly below 180 degrees, see Table~\ref{Models}).

\begin{figure}
	\includegraphics[width=\columnwidth]{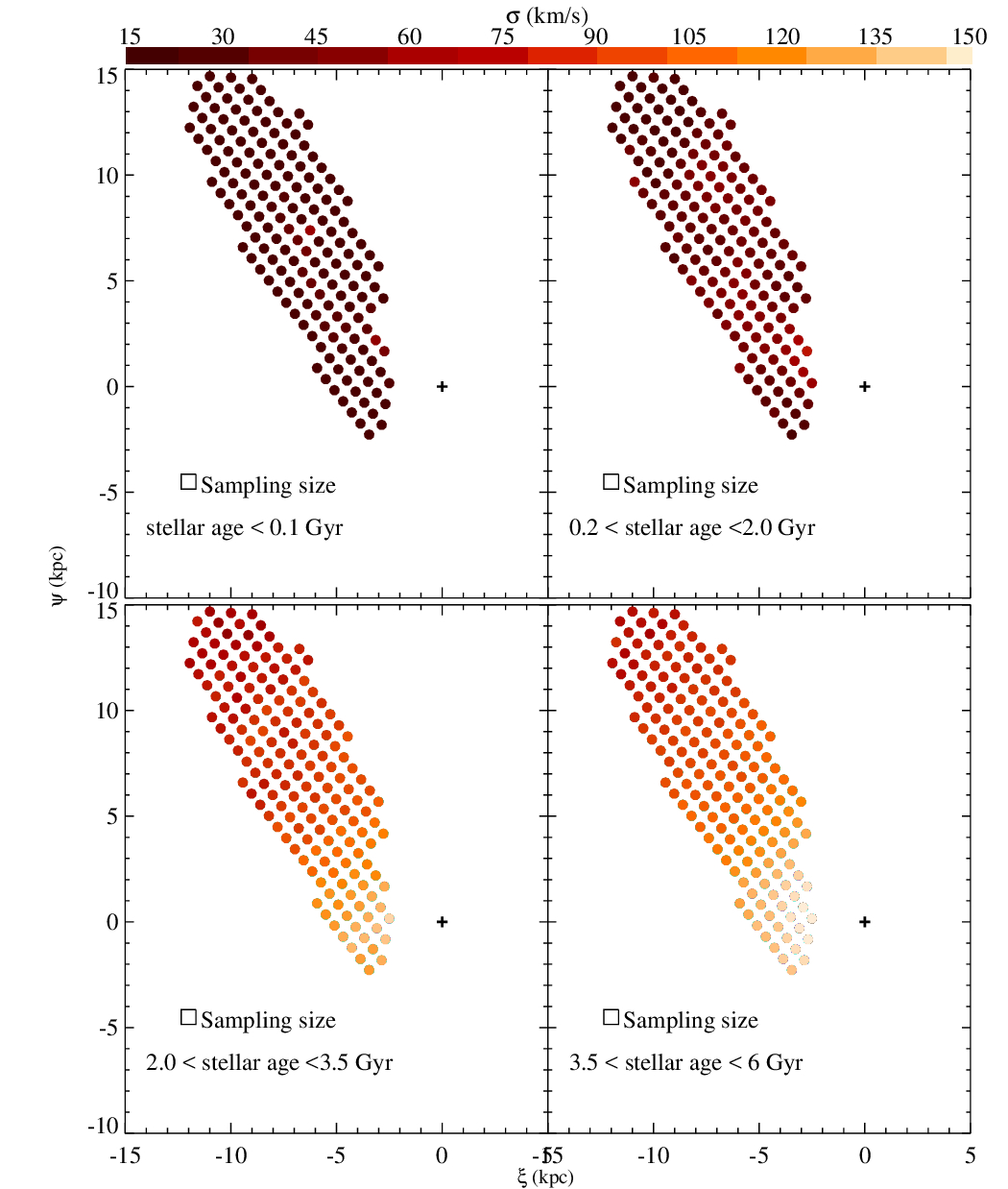}
    \caption{Maps of the velocity dispersions of stars of different ages for a direct comparison to Fig. 7 of \citealt{Dorman2015}. Since 20 M particles are necessary to achieve sufficient particle number per pixel, we show only the results for Model 336. The cross represents the M31 centre.}
    \label{starkinmap}
\end{figure}

A slight difference between our model estimates and the observations in Figure~\ref{starkin} comes from the fact that we lack a sufficient  number of particles outside the projected major axis of the M31 disk. To improve the statistics, we run a 20 M particles simulation, Model 336, for which the parameters are exactly those of Model 314, except for a 5 degree offset on the angular position of the secondary and an increase of the progenitor gas fractions (see Table~\ref{Models}). Figure~\ref{starkinmap} shows the 2D dispersion maps of the modelled M31 assuming 4 age ranges, which correspond to main sequence, young and old AGB and RGB stars, from the best expectations based on Fig. 9 and 12 of \citealt{Dorman2015}. It reveals a remarkably similar $\sigma_v$ map to the observed one (see Fig. 7 of \citealt{Dorman2015}), both quantitatively and qualitatively. The age-velocity dispersion relation and its steepness is widespread over all the disk, and the velocity dispersion increases significantly at low radii. Figures~\ref{starkin} and ~\ref{starkinmap} show that our modelling also reproduces the dispersion versus radius relationship shown in Fig. 16 of \citealt{Dorman2015}, i.e., a strong increase of the dispersion below $R = 10$~kpc. As with the observations, our modelled 2D maps show local variations in the velocity dispersion. It is beyond the scope of the present contribution to acquire sufficiently detailed maps for, e.g., retrieving the influence of the bar as proposed and tested by \citet{Dorman2015}.

\subsection{The unusual properties of the M31 outskirts: Giant Stream, shells, clumps and warps}
\label{outskirts}
The main success of minor merger models \citep{Fardal2007,Fardal2008,Fardal2013} is to reproduce simultaneously the NE and W shells together with the Giant Stream (GS). The association with a less than 1 Gyr old event is also consistent with the sharpness of the observed shells and Giant Stream, which are dynamically ``young'' features. Shells are also formed in major mergers as shown by observations and by, e.g., the modelling of NGC 7600 by \citet{Cooper2011}. However the major merger model of \citet{Hammer2010} did not reproduce well these features, which might be due to the small particle number (200k for stars) used in that study. However, a  coalescence of the nuclei occurring 5.5 Gyr ago let few chances to recover the sharpness and contrast of the observed features. We have verified that such an ``ancient'' merger in M31 would underestimate the amplitude of the velocity dispersion in the M31 disk by a significant factor ($\sim$ 2).

\begin{figure*}
	\includegraphics[width=2\columnwidth]{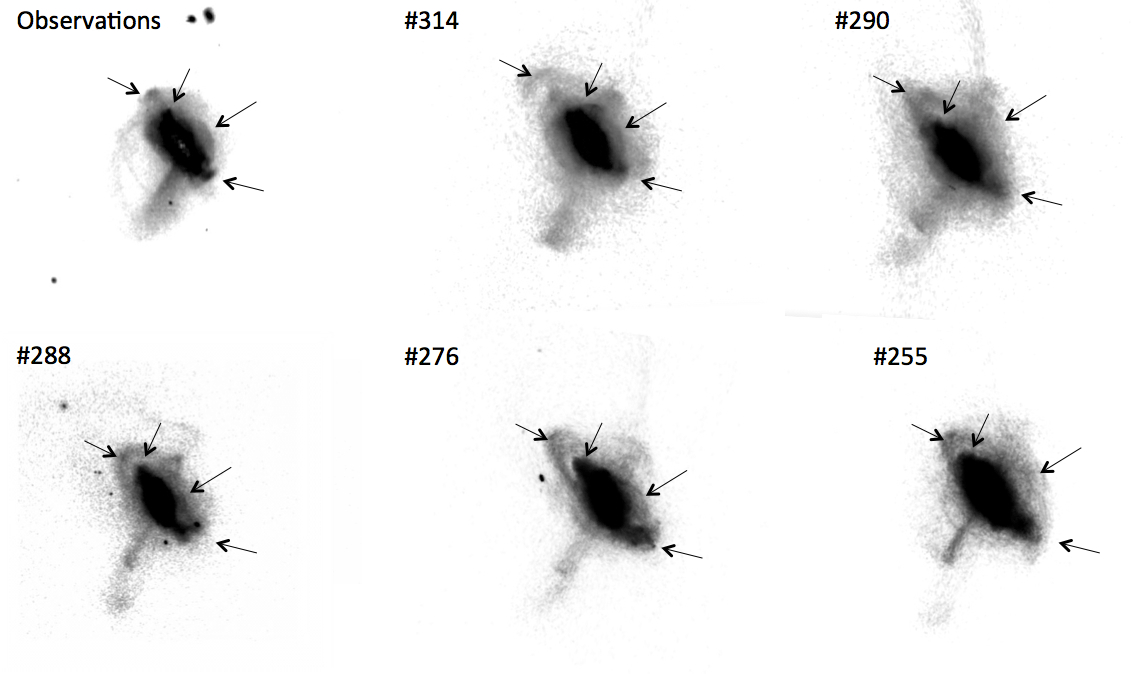}
    \caption{A comparison of the PAndAS imaging (first panel on the top-left) with that of five selected models. The height of each panel is 264 kpc. The arrows indicate the most obvious features besides the GS, from left to right, top to bottom:  NE Clump, N Spur, W shell, and G1 clump. Notice that the resolution is $\sim$ 16 times better in the observations than in the models.}
    \label{5models}
\end{figure*}

Since a 2--3 Gyr old major merger is an excellent contender for explaining the kinematics and star formation of the M31 disk, one may wonder if this event could be sufficiently recent to be also responsible for the GS, the NE and W shells, including their sharpnesses. Figure~\ref{5models} shows that the fiducial models lead to sharp GS-like structures and shells, though the NE shell is often smaller than in the observations except perhaps for model 314 for which it seems even too prominent. The models show a strong warp on both sides of the disk main axis, which share many similarities with the observed Warp and Northern Spur regions.  Table~\ref{SBTab} gives surface brightness ratios between several features and the GS that range around the observed values within a factor $\sim$ 2. Moreover, Figure~\ref{5models} shows that the NE and G Clumps are more prominent than those observed. We note that changing the progenitor sizes provides different Clump morphologies and a better reproduction of them would require a better knowledge of the initial gas and stellar spatial distribution in both progenitors. The NE and G Clumps have been posited to be tidal debris or undergoing tidal disruption \citep{Zucker2004, Ibata2005} from their disorderly morphologies, shared by their gaseous counterparts \citep{Lewis2013}.  In our modelling, both clumps correspond to tidal dwarfs associated with the main progenitor that are in a process of disruption, and their kinematics and stellar content are likely similar to those of the M31 disk, which would reconcile former interpretations of the G1 and NE Clumps.

\begin{table}
	\centering
	\caption{Surface brightness ratios between the following fields: F1 (21 kpc, halo field of \citealt{Brown2006}), F2 (Giant Stream field of \citealt{Brown2007}), F3 (warp field), F4 (North-East shelf), the F2 field has been taken as the reference.}
	\begin{tabular}{lcccc} 
		\hline
		 & F1 & F2 & F3 & F4\\
		\hline
		Obs   &  0.08  & 1.00  & 3.23 &  0.91\\
		\hline
	Model  & & & &\\
		255   &  0.11  & 1.00 &  5.06  & 0.46\\		
		276   &  0.10  & 1.00  & 7.66  & 0.67\\
		288  &   0.17  & 1.00  & 9.79  & 3.10\\
		290   &  0.15  & 1.00 &  2.76 &  0.95\\
		314   &  0.02  & 1.00  & 6.98  & 1.53\\
		\hline
	\end{tabular}
  \label{SBTab}
\end{table}

In Figure~\ref{5models} the GS is mostly the superposition of the first and second loops associated with the tidal tail linked to the second passage of the secondary (see Sect.~\ref{scenarios}). The loop plane is seen edge-on since the first and second tidal tails share the same plane as the disk of satellites, explaining then why it points to the MW (see, e.g., \citealt{Hammer2013}). The second loop is the most recent and the brightest one and could explain the sharpness of the GS structure. Since the brightest (second) loop is further away than M31, this is in qualitative agreement with the distance behaviour of the GS described by \citet{Conn2016} (see their Fig. 4), and a more detailed description can be found in Sect.~\ref{scenarios}. Finally the GS shape strongly depends on the initial orientation of the secondary, though we consider that the contrast and sharpness of the GS have been largely recovered by our modelling.

\begin{figure}
	\includegraphics[width=\columnwidth]{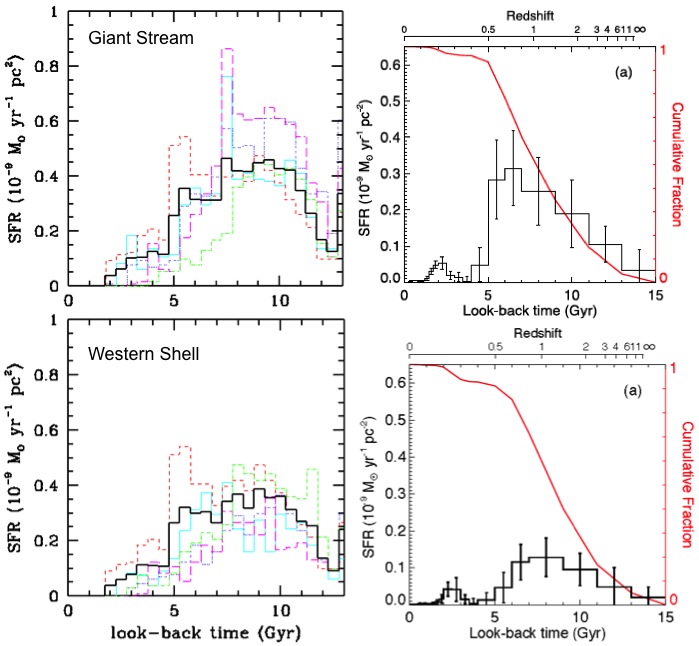}
    \caption{{\it Top row:} Star formation history derived for the Giant Stream, on the left based on the five models shown in Fig.~\ref{5models} (thick black line indicates the average), on the right as it has been measured by \citealt{Bernard2015a}. Cyan, blue, red, magenta, and green represent models as in Fig.~\ref{DISK_ages}. {\it Bottom row:} Same as above for the star formation history of the W shell, on the right as it has been measured by \citealt{Bernard2015b}.}
    \label{SFHs}
\end{figure}

Figure~\ref{SFHs} shows the stellar age distribution in both the GS and the W shell. It illustrates well how the modelling is able to reproduce the GS age distribution as well as the SFR density amplitude. As it is observed, the age distribution of the W shell is similar to that of the GS. In fact both structures are linked to the material of the secondary encounter that dominates the GS and the shell. In Figure~\ref{decomp} we decompose the particles of the primary (left panel) and the secondary (middle panel) progenitors. This suggests that the substructures in the outer M31 disk are of tidal origin, with warps and clumps being linked to the primary, while the GS and the shells are mostly the imprints of particles of the secondary progenitor.

\citet{Bernard2015b} have separated outer disk structures into disk-like (warps and clumps) and stream-like (GS and shells) categories, respectively, depending on their star formation history and metallicity properties. Figure~\ref{decomp} shows that a major merger model can produce precisely these two kinds of structures. Notice, however, that a few particles of the primary have been captured by the secondary during the interaction and can be found in the GS as well as in the shells.



\section{Discussion and conclusion}
\subsection{Limitations, weaknesses, and successes of a major merger modelling of M31} 
In this paper, we have investigated the properties of a potential major merger that may have occurred in M31. The main challenges arise from the gigantic number of observational constraints provided by the numerous surveys that have scrutinised our closest neighbour, and the huge contrast between the brightest and faintest features (e.g., a factor of 1000 between the central bulge and the GS). Furthermore, the uncertainties linked to the pericentre determination (at best within 4\%) could change by 1 Gyr the elapsed time between the 1st and 2nd passage, limiting significantly our capability to optimise the modelling. Large pericentre values are however mandatory to rebuild disks as large as that of M31. To mitigate the uncertainties, we have chosen a pericentre radius of $r_p$ $\sim$ 32 kpc, which has led us to mr$\sim$ 4. We conjecture that a smaller mass ratio (e.g., 3:1) would require a higher pericentre to provide similar results. On the other hand, larger mass ratios (e.g., mr$\ge$ 4.5) are excluded since they cannot produce sufficiently large disks in the remnant galaxy. Our current modelling leads us to assume $r_p$$\ge$ 32 kpc and mr$\le$ 4.25 as robust limits.

\begin{figure}
	\includegraphics[width=\columnwidth]{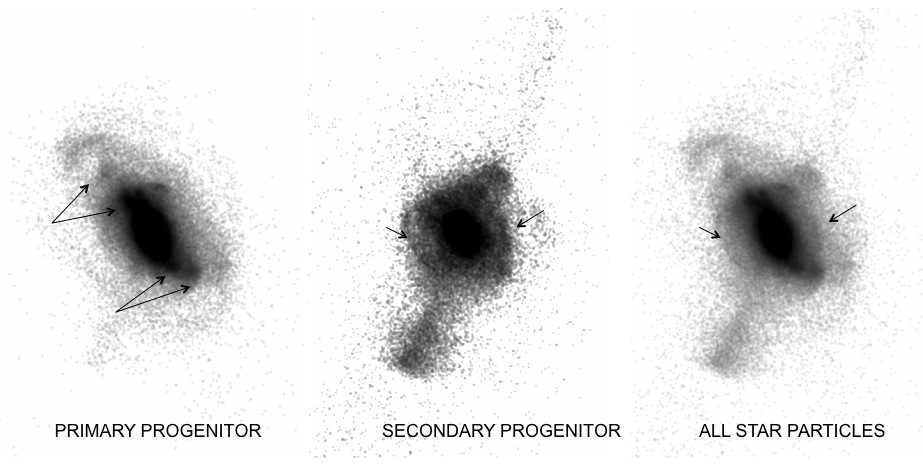}
	\includegraphics[width=\columnwidth]{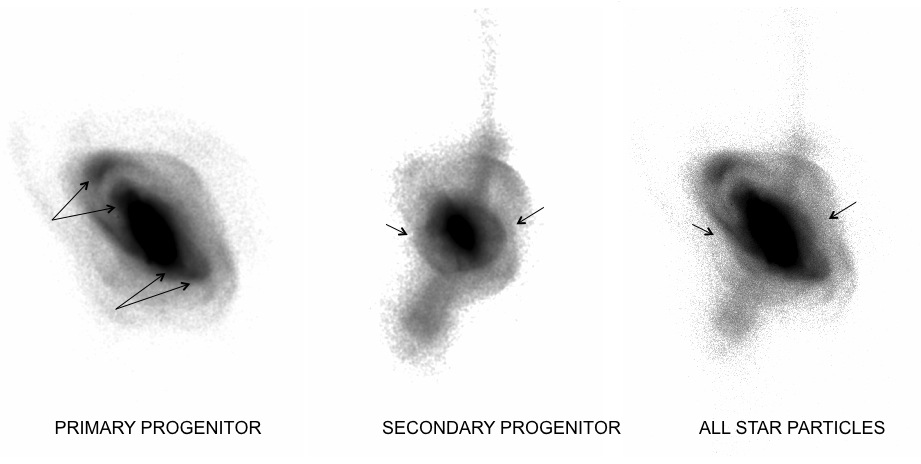}
    \caption{Decomposition of stellar particles from their origin into the two progenitors in model 314 (Top row) and in model 336 with 20 M particles (Bottom row). The height of each panel is 264 kpc. {\it Left:} Particles from the main progenitor. The arrows indicate the warps and the clumps. {\it Middle:} Stellar particles from the secondary. The arrows indicate the shells. {\it Right:} All stellar particles. The arrows indicate the shells.}
    \label{decomp}
\end{figure}

Despite these limitations, the modelling of M31 as a recent (major) merger is able to reproduce most internal properties of M31 (bulge, bar, 10 kpc ring, stellar and \ion{H}{I} disks) including its rotation curve and disk dispersion. This has been done after an intensive search to optimise the initial orbital parameters, and those linked to the progenitor structures (see Table~\ref{Models}). Disk-like features (warps and clumps) associated to the outer disk are also produced by the modelling as well as stream-like features such as the GS and shells (see Figures~\ref{5models} and ~\ref{decomp}).  Figures~\ref{DISK_ages} and~\ref{SFHs} confirm the ability of the model to reproduce the star formation histories of these structures. 
 
 This apparently contradicts the conjecture made by \citet{Hammer2010} that the $\ge$ 5 Gyr old star population of the GS can be used as a clock, implying a similarly old  coalescence epoch. While this is probably true for their adopted 3:1 mass ratio and 25 kpc pericentre, by increasing mass ratio (to 4:1) and pericentre (to 32 kpc), we have somewhat unsettled the clock. The GS is made of particles returning to the remnant from the tidal tail formed at the second passage, mostly including material from the secondary (see middle panels in Figure~\ref{decomp}). First, as soon as the material is deposited on the tidal tail the star formation ceases, and the time elapsed between the second passage and the  coalescence of the nuclei (as well as the number of passages) is increasing with the mass ratio, due to a decreasing dynamical friction. Second, one or two Gyr before the second passage, the star formation in the secondary is likely smaller for large pericentres, because the system has a much larger time to relax. In this paper we have verified  that a 4:1 merger, 32 kpc pericentre merger reproduces together the 2-4 Gyr old burst of star formation, the GS star formation history and the age-dispersion relation in the disk.  
 
\begin{figure}
	\includegraphics[width=\columnwidth]{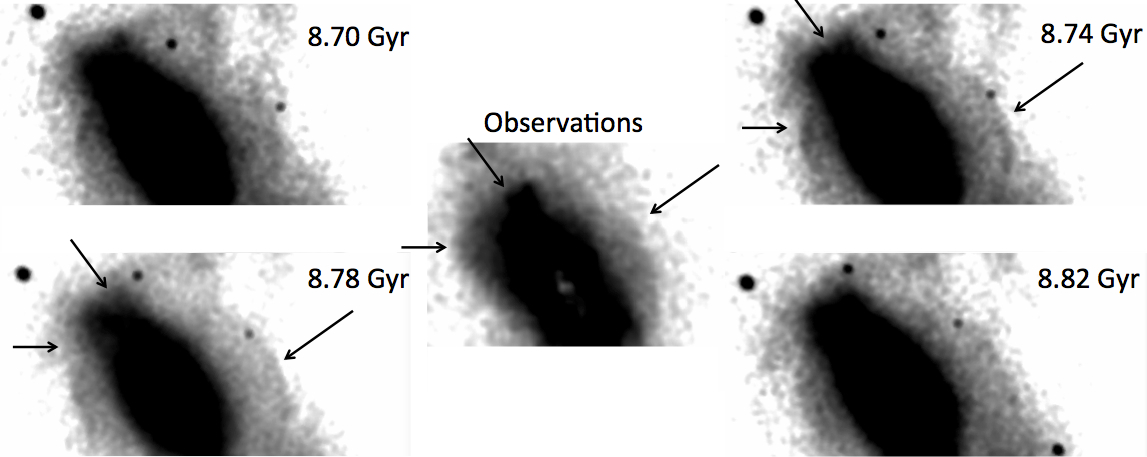}
    \caption{{\it Central image:} Zoom on the thick disk outskirts from PAndAS \citealt{Ibata2014}.{\it Outer panels from top-left to bottom-right:} Evolution of the simulations (Model 288) over 120 Myr, showing that many substructures near the thick disk are evolving rapidly, including shells (see arrows).}
    \label{evolution_thickdisk}
\end{figure}

However, Figure~\ref{5models} does not provide an accurate reproduction of the observed outer disk features (see also the right panel of Figure~\ref{decomp}).  For example, the simulated NE Clump is often more extended than what is observed. There is also a more prominent counterpart on the opposite side of the GS than in the observations.  It appears that the former discrepancy depends on the precise morphology of the progenitors, while the later is reduced by increasing the mass ratio and the initial gas fraction. There is another discrepancy that considerably alters our modelling ability to reproduce features, altogether. This is caused by the occurrence of shells that seem to be highly variable with time, restricting the possibility to accurately model the outskirts of the M31 disk. The W and NE shells of M31 seem very similar to shells seen in simulations of elliptical galaxies (see, e.g., \citealt{Cooper2011}), i.e., shells are arranged along the orbital path of the secondary, to which they are linked, as demonstrated by Figure~\ref{decomp} (see middle panels). Our modelling shows that shells are as variable as those seen in the \citet{Cooper2011} simulations, and Figure~\ref{evolution_thickdisk} demonstrates that the variation time-scale may reach values below 100 Myr. 
  
\subsection{A single major merger or a multiple minor merger paradigm?} 
\label{scenarios}
\begin{figure}
	\includegraphics[width=\columnwidth]{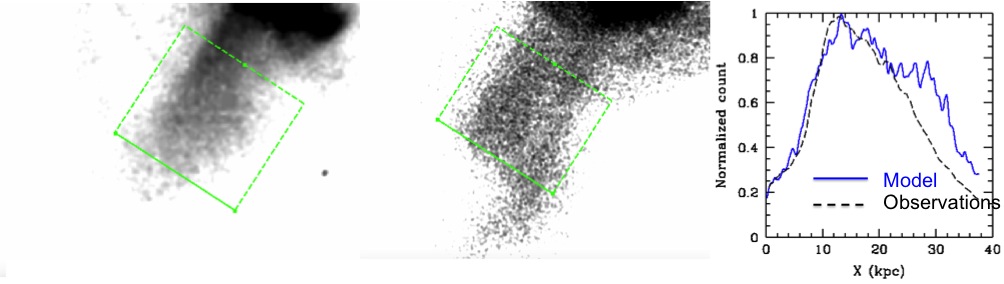}
    \caption{{\it Left:} Zoom on the GS from PAndAS (see \citealt{Ibata2014}), the green box indicating how the profile across the GS has been derived.{\it Middle:} Same for Model 336.{\it Right:} Profile across the GS. }
    \label{GS}
\end{figure}

The numerous studies of the GS \citep{Fardal2006,Fardal2007,Fardal2008,Fardal2013,Mori2008,Kirihara2014,Kirihara2017, Sadoun2014} in a minor merger context have been very successful in reproducing the detailed features of the stream such as, e.g., the asymmetric profile across the GS \citep{Kirihara2017}. Figure~\ref{GS} shows that such a property can be retrieved also by our modelling. Interestingly, we have verified that orbits taken for the minor merger (e.g., from \citealt{Fardal2006}) share similarities with our orbital parameters, especially for the second passage that determines the spatial structure of the GS. This suggests that a major merger (as well as a minor) is able to reproduce the GS together with NE and W shells, those structures being mostly attached to the secondary or satellite (see middle panels of Figure~\ref{decomp}), while other structures such as warps and clumps are mostly attached to the primary progenitor (see left panels of Figure~\ref{decomp}). 

In minor merger models, satellite (or secondary) masses range from 3-4 $10^{9}$ \citep{Fardal2013,Kirihara2014} to 4.2 $10^{10}$ $M_{\odot}$ \citep{Sadoun2014}. It suggests that the success in modelling the GS depends more on orbital parameters than on mass ratio. \citet{Sadoun2014} reduced the GS-compatible mass ratio by a factor $\sim$ 15 (from 300:1 to 21:1), and the present work allows it to be extended by a supplementary factor 5, to 4:1. We have verified that 2:1 to 5:1 models (see Table~\ref{tbICs}) may also reproduce the same features.

 The main theoretical support for a minor merger explanation for the GS and other streams comes from the simulations of \citet{Bullock2005}, which predicted a considerable number of minor accretion events. This influential study predicted that mass deposited by hundreds of satellites could explain the stellar content of galactic haloes and their associated streams. However, the prediction of hundreds of sub-haloes, e.g., for the MW or M31, is not observed, dubbed as the "missing satellites" problem (see, e.g., \citealt{Bullock2017}, or for an alternate viewpoint, \citealt{DelPopolo2017}). Perhaps the evidence for multiple minor mergers having occurred in M31 has to be reconsidered. In fact, semi-empirical $\Lambda$CDM models have been proposed by \citet{Hopkins2010} and systematically compared to observations. It emerges that at $z=0$ and for a $M_{stellar} \ge$  $10^{11}$ $M_{\odot}$ galaxy (such as M31), the occurrences in the past 10 Gyr of a 4:1, 21:1 and 300:1 merger are  70, 108 and 160\%, respectively. In other words, a major merger scenario for M31 would have a similar occurrence as a scenario with $\sim$ 2 satellites, instead of 15 if each stream is linked to a single event \citep{Tanaka2010}. 

\begin{figure}
	\includegraphics[width=\columnwidth]{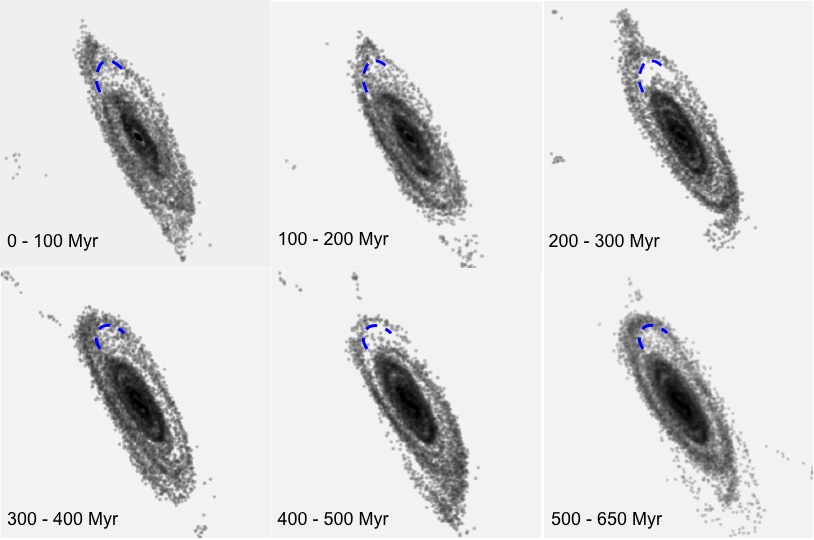}
    \caption{Maps of the SFH density ($M_{\odot}yr^{-1}kpc{-2}$ of model 336 using 650  to 0 Myr old stars. The time range covered is shown in the lower left of each plot, and the Figure can be directly compared to the observations provided in Fig. 5 of \citealt{Lewis2015}, including the blue dashed curve that aids the eye in recognising structural change between time bins. The size of each panel is 40 kpc.}
    \label{ring_evolution}
\end{figure}

\begin{figure}
	\includegraphics[width=\columnwidth]{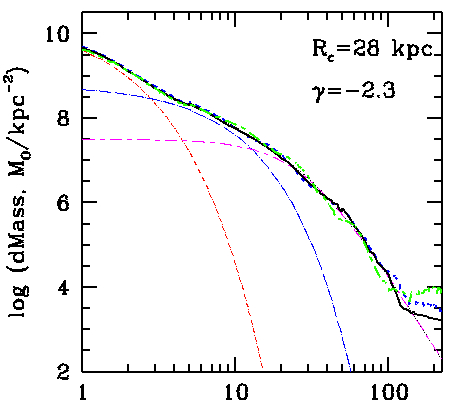}
    \caption{Stellar halo profile for mr=4 (thick black line, average of the 5 models of Table~\ref{Models}), mr=3 (thick dot, blue line) and mr=5 (thick, short dash - long dash, green line). Models of the bulge, the disk and the halo are provided by thin lines, dot - short dash (red), dot - long dash (blue), and short dash - long dash (magenta) lines, respectively. The adopted halo profile follows a cored power-law (1 +(R/$R_c)^2)^{\gamma}$), as in \citealt{Ibata2014}. The fit of the halo with $\gamma$=-2.1 to -2.3 is valid until  100, 120, and 130 kpc for models with mr=5, 4 and 3, respectively.  Note that the progenitors in our simulations have no initial stellar halos.}
    \label{halo}
\end{figure}

A major merger paradigm is supported by the widespread  age-dispersion relation in the disk of M31. In fact, a 21:1 merger cannot provide the velocity dispersion amplitude observed in stars lying in the overall M31 disk: for example, at $R= 10$ kpc, Fig. 20 of \citet{Sadoun2014} predicts $\sim$ 30 km$s^{-1}$ that compares to 80-100 km$s^{-1}$ observed for the bulk of stars by \citet{Dorman2015}. It is recovered by our models (see Figures~\ref{starkin} and ~\ref{starkinmap}), due to the highly retrograde orbit of the primary adopted in this paper (see Table~\ref{Models}), which is particularly efficient in stressing the primary disk. We verified that $\ge$ 5:1 mass ratios are not sufficient to reproduce the observed velocity-dispersion amplitudes.    
A recent major merger is also consistent with important properties of M31, which are not successfully explained for the moment:
\begin{enumerate}
\item  The 10 kpc ring that is a long-lived feature that has survived for at least 400 Myr \citep{Davidge2012,Dalcanton2012,Lewis2015}; Figure~\ref{ring_evolution} shows that in a 4:1 post-merger, the ring is naturally stable for at least 500 Myr and our modelling also captures the presence of the outer ring at 15 kpc (see Figure~\ref{barring}).
\item The 2--4 Gyr star formation event that is widespread over the whole M31 disk (see, e.g., discussion in \citealt{Williams2015}) and consistent with a recent 4:1 merger (see Figures~\ref{DISK_ages} and~\ref{DISK_cumul_ages}).
\item The NE and G Clumps are predicted together with a warp in a major merger; \citet{Ferguson2016} recalled that these structures have been supposedly linked to the disruption of a dwarf, which is the case in our modelling (more precisely a tidal dwarf), together with the fact that they have disk-like properties such as the warps.
\item The halo profile predicted by a 3--4:1 major merger (see Figure~\ref{halo}) is consistent with the observations \citep{Gilbert2012,Ibata2014}, with a constant (projected) slope of $\gamma$=-2.1 to -2.3, and without a drop until 120 kpc. This property is shared by modern cosmological simulations including major mergers (see, e.g., Fig. 4 of \citealt{Font2011}); in contrast and as noticed by \citet{Gilbert2012}, ten among eleven simulations of the dwarf dominated simulations by \citet{Bullock2005} show a break to a steeper profile before or at 100 kpc.
\item The absence of evidence for a residual core of the satellite responsible of the GS formation is expected since all the secondary material has been wiped out into the different structures of the remnant (see Figure~\ref{decomp}).
\end{enumerate}

\begin{figure*}
	\includegraphics[width=1.8\columnwidth]{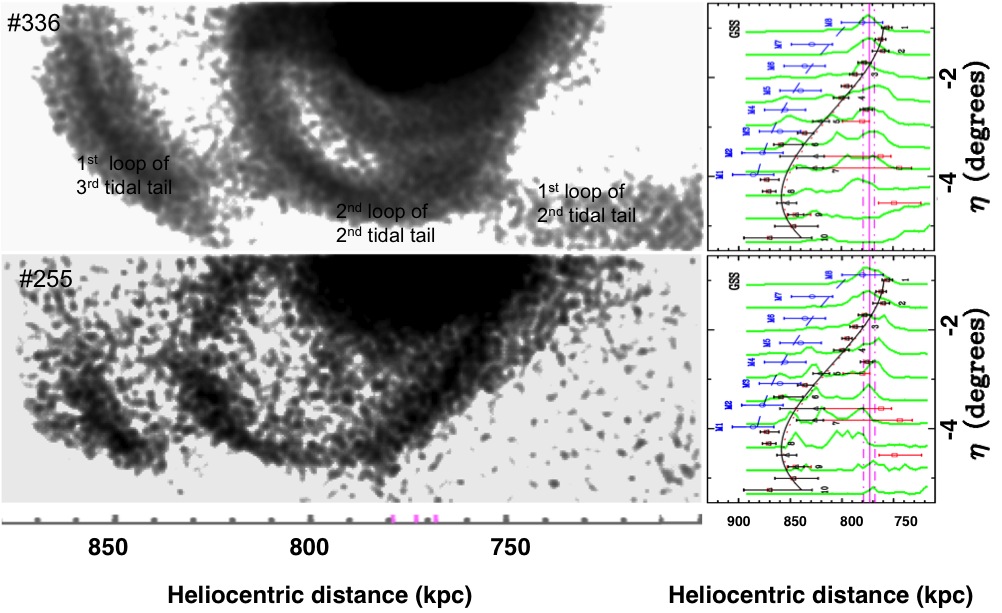}
    \caption{3D structure of the GS for Models 336 (top, 20M particles) and 255 (bottom, 2M particles). The centre of M31 is taken at 773 kpc from the MW (see magenta tick marks and vertical lines in left and right panels, respectively) as in \citealt{Conn2016}. {\it Left:} Zoom on the GS components (or loops) for model 336 in a (Z, Y) plane. Z corresponds to distance from the MW (from left to right: 873 to 673 kpc), and Y is the distance in kpc from the M31 centre in the direction from North to South (from top to down: 20 to 100 kpc). {\it Right:} Reproduction of Fig. 4 of  \citealt{Conn2016} for which  the solid line represents the best estimate of the GS distance distribution, and red squares correspond to alternatives due to multiple peaks. Modelled mass profiles (green lines, extracted from horizontal cuts of the left image) have been superposed on the observations. Although the simulations do not reproduce accurately the observations, they do show a complexity that is quite similar. As in the observations they show a single peak at the top, then multiple peaks near $-3$ degrees and below.  Models also show that from the top to bottom, distances are increasing and then slightly decrease, as in the observations.} 
        \label{GS_dist}
\end{figure*}

 The 3D structure of the GS may provide a further, and possibly definitive test of the GS origin.  \citet{Conn2016}  measure the distance from the MW of stars along the GS. It reveals some complexity from the mid part to the end of the GS, for which they found probability distributions  that are often double peaked. Figure~\ref{GS_dist} suggests that these multiple peaks are real and are caused by a complex 3D distribution as is expected if the GS is the superposition of 2 or even 3 loops (see left panels).  If correct, this causes an increasing number of different density peaks with distance from M31, in qualitative agreement with what is observed. Such complexity may be difficult to explain if the GS is a single tidal tail caused by the passage of a satellite (see, e.g., Fig. 8 of \citealt{Kirihara2017}). Notice that contamination by other components is unlikely in fields this far from the M31 thick disk.

\subsection{Concluding remarks}
Galaxies are made by  mergers of smaller entities according to the hierarchical scenario, for which M31 is perhaps one of the best illustrations. The relative success of a major merger for explaining most of the large-scale properties of M31 does not mean that we have securely identified its past history and fully determined the parameter set of the former interaction. We reach robust, though not precise, predictions for the mass ratio ($<$4.5:1) and for the pericentre ($\ge$ 30 kpc), in order to build sufficiently large stellar and \ion{H}{I} disks. This leads to relatively precise predictions for  the coalescence of the nuclei epoch (1.8 to 3 Gyr ago) to reproduce the age-dispersion relation and the 2--4 Gyr old star formation event, and for the angular orbital parameters, which lie within a small range of values to reproduce the 10 kpc ring and the GS (see Table~\ref{Models}). This does not preclude of any other, additional minor merger or interaction, such as that with M33 (see, e.g., \citealt{McConnachie2009,McConnachie2010}) creating a faint and shallow bridge between the two galaxies. If correct, the two events would have happened at a similar epoch (few Gyr ago), and a fly-by with M33 is likely less influential than a major merger  at the time when the two nuclei have coalesced.

 An important conclusion of this paper is that the GS and shell modelling depends sensitively on angular orbital parameters, but is almost (or fully) mass-ratio independent, i.e., they can be produced either by a single minor or major event. 
Furthermore, a recent major merger appears to be a rather compelling scenario to explain the disk age-dispersion relation, and to reproduce a realistic 10 kpc ring, the 2--4 Gyr old star formation event, the NE and G clump properties, the halo profile, and perhaps the 3D structure of the GS.

\begin{figure}
	\includegraphics[width=\columnwidth]{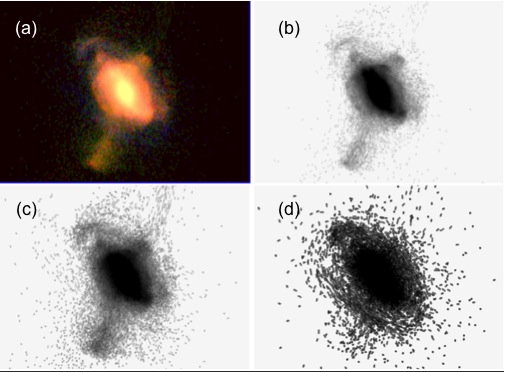}
	\includegraphics[width=\columnwidth]{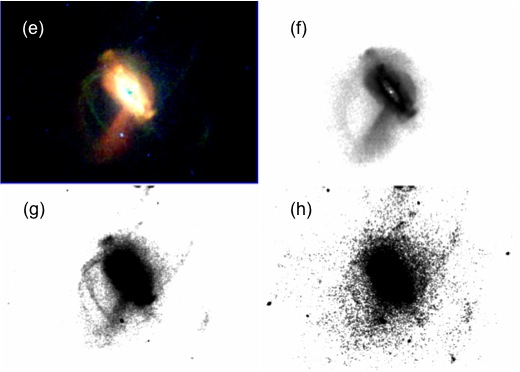}
    \caption{Metal distribution from Model 314 (top 4 panels, a, b, c, d) and that observed in M31 (bottom 4 panels, e, f, g, h), for which the resolution is 16 times higher than the former. {\it Top left:}) RGB colour combination of stars with log($Z/Z_{\odot}$) in the range of -2.5 to -1.8 (blue), -1.8 to -1 (green), and -1 to 0 (red), respectively (panel a: modelling; panel e: observations). In the modelling, panels b, c and d (observations: panels f, g, h) represent the corresponding distribution of low (blue), intermediate (green), and high (red) metallicity (coded stars), in the respectively. In the progenitors we have estimated metallicity assuming it decreases radially  by -0.1 dex/kpc, similar to expectations in the MW, 7 Gyr ago (see, e.g., \citealt{Prantzos2007}). Initial average values for each progenitor have been taken from \citealt{Rodrigues2012}, assuming the observed slope of the mass-metallicity relation at z$\sim$ 1.5 (9 Gyr ago). }
    \label{metal_decomp}
\end{figure}

The outer disk structures of M31 have bimodal age and metallicity properties: GS and shells on one side, disk-like structures such as warp and clumps on the other side. This is precisely mirrors the material that originated in the secondary and the primary, respectively (see Figure~\ref{decomp}). Figure~\ref{metal_decomp} compares an rgb decomposition of the metal abundances for model 314 with what is observed in M31 \citep{Ibata2014}. They both show an increase of star number density with metallicity (compare b and d panels and then f and h panels), and a quasi absence of low metallicity stars at the location of the GS. M31 has been found to have a more spherical halo at low metallicity \citep{Ibata2014}, consistent with these simulations.

Intermediate and high metallicity stars of model 314 show a hint for a possible bifurcation at the GS end, which might suggest stream B and C. However the simulations would require $\sim$ 16 times more particles to match the observations. To overcome this difficulty would require us to compute a significant number of 20M particle simulations, which is far beyond the scope of this study, since it would need a significant increase in calculation speed. In this paper, we are left to speculate that stream B (and perhaps stream C) might have originated in the secondary progenitor, while stream D might belong to the very outer tidal tail of the primary (see and compare low-metallicity panels d and h in Figure~\ref{metal_decomp}).

Predictions or tests are often useful to support (or to dismiss) a scenario. If M31 has been structured after a recent, major merger responsible for the formation of the GS, we find that returning stars from the second tidal tail should be at least, barely observable at the depth of the PAndAS survey. Our modelling then predicts a wide and ultra faint over-density of stars to the North of the M31 halo, in the region including And XXVII, XXV, XXVI and perhaps also NGC 147 and 185 (see, e.g., Fig. 1 of \citealt{Richardson2011}). In Figure~\ref{halo} one may notice, at large radii ($>$ 120 kpc), a flattening of the profile instead of a steepening. Such a flattening is due to the recent  coalescence of the nuclei since it vanishes at later times. We then predict an overabundance of (especially intermediate and metal rich) stars at 130 to 200 kpc projected distance from M31, on both Northern and Southern sides, while we would expect an under-abundance on the Western side.

Finally, one may notice that tidal tails attached to the secondary lie in a single thin plane, which is the orbital plane. In all models shown in Table~\ref{Models}, 
it includes the MW, i.e., matching very well the plane of satellites found by \cite{Ibata2013}. 
This result seems robust and linked to our viewpoint at the MW location, which requires us to see the GS as the superposition of loops, i.e., almost edge-on. 
Such a link may lead to clues about the origin of vast planar structures of satellites, which are 
not fully understood yet \citep{Pawlowski2014}. Could their origin be linked to an
 interaction between two satellite systems during the merging, and under which conditions they can lead to such a thin structure as that surrounding M31 \citep{Smith2016}? If the thinness of the plane of satellites is linked to the thinness of the merger orbital plane, it could even lead one to question the nature of M31 dwarfs, as suggested by \citet{Hammer2013}.  In our modelling, the G1 and NE clumps are disrupted tidal dwarfs (or fluctuations in a tidal tail), and they would be linked with tidal tails attached to the primary. The latter formed at the same epoch as the (second) tidal tail in the orbital plane, i.e., $\sim$ 1 Gyr before the  coalescence of the nuclei, which would not give much predictive power to comparisons between dwarfs in and out the plane \citep{Collins2015,Collins2017}. Future simulations with large particle numbers have to be done, including, e.g., interactions between pre-existing satellites and/or tidal dwarf systems, to verify under which conditions the former may provide thin planes of satellites and/or the later may provide dwarfs with stars as old as that in the M31 dwarfs, knowing that part of their tidal material is common with that of the GS (see Figure~\ref{SFHs}).
 
\section{Acknowledgements}

This work was granted access to the HPC resources of TGCC/CINES/IDRIS under the allocation 2016-(i2016047633) made by GENCI, and to MesoPSL financed by the "Region Ile de France" and the project Equip$@$Meso (reference ANR-10-EQPX-29-01) of the program "Investissements d'Avenir" supervised by the "Agence Nationale de la Recherche". This work has been supported by the China-France International Associated Laboratory "Origins". J. L. W. thanks the China Scholarship Council (NO.201604910336) for the financial support. We are grateful to Phil Hopkins who kindly shared with us the access to the GIZMO code. F.H. warmly thank Lia Athanassoula for her precious and wise advices on modelling  substructures of an individual galaxy.







\appendix
\section{Mass profile and rotation curves}
\begin{figure*}
	\includegraphics[width=1.5\columnwidth]{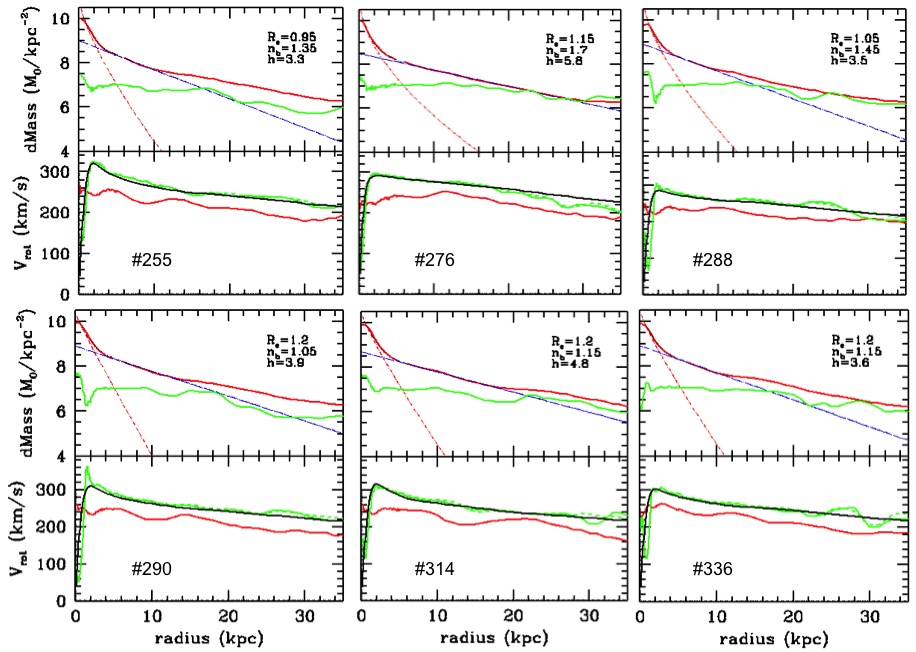}
    \caption{Mass profile and rotation curve for 6 models (see the model number on each bottom panel) up to 35 kpc from the centre. {\it Top panels:} stellar mass profile (red) that is fitted by 2 components (red-dashed line: bulge with a free Sersic index, blue long-dashed line: exponential disk). Green line indicates the gas profile. On the top-right of each panel, the bulge effective radius and Sersic index, as well as the disk scale-length, are given. {\it Bottom panels:} Rotation curve for each component (red: stars, green: gas) and from the virial theorem (black curve). Notice that the gas is sufficiently relaxed after the 1.8-3 Gyr old merger to follow well the theoretical curve,. It also matches quite well the observed rotation curve by \citealt{Chemin2009}. }
   \label{RC_PRO}
\end{figure*}

Figure~\ref{RC_PRO} shows the stellar mass profiles (top panels) and the rotation curves (bottom panels) for the 6 models tabulated in Table~\ref{Models}. Based on these fits, B/T ratios are ranging from 0.35 to 0.42, which is consistent with observations after assuming that B accounts for both bar and bulge.

\section{Convergence of simulations}
\begin{figure*}
	\includegraphics[width=1.5\columnwidth]{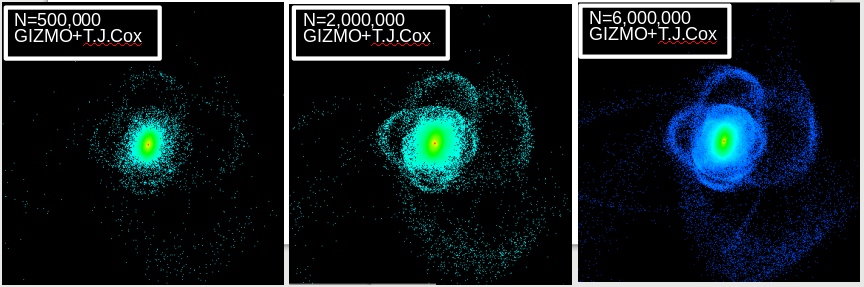}
    \caption{From left to right: result of a simulation of a 3:1 gas-rich merger showing loops and tidal tails in the orbital plane with 500 k, 2M and 6M particles, respectively. Pericentre is 25 kpc, and initial gas fractions are 50 and 80\% in primary and secondary progenitors, respectively.}
   \label{converge}
\end{figure*}

Figure~\ref{converge} presents the results of the same simulation of a 3:1 merger at 3 different resolutions. Snapshots have been captured 7.2 Gyr, 6.2 Gyr, and 3.5 Gyr after the beginning of the simulation, the first passage and the  coalescence of the nuclei, respectively. Notice that most details in the loops and tails are well preserved after passing from 500k to 6 M particles (see Wang et al., in preparation).



\bsp	
\label{lastpage}
\end{document}